Article

# A framework to determine micro-level population figures using spatially disaggregated population estimates

V.E Irekponor*[1] M. Abdul-Rahman[1,2] M. Agunbiade[1] A.J Bustamente[3]

**Abstract**

About half of the world population already live in urban areas. It is projected that by 2050, approximately 70% of the world population will live in cities. In addition to this, most developing countries do not have reliable population census figures, and periodic population censuses are extremely resource expensive. In Africa's most populous country, Nigeria, for instance, the last decennial census was conducted in 2006. The relevance of near-accurate population figures at the local levels cannot be overemphasized for a broad range of applications by government agencies and non-governmental organizations, including the planning and delivery of services, estimating populations at risk of hazards or infectious diseases, and disaster relief operations. Using GRID3 (Geo-Referenced Infrastructure and Demographic Data for Development) high-resolution spatially disaggregated population data estimates, this study proposed a framework for aggregating population figures at micro levels within a larger geographic jurisdiction. Python, QGIS, and machine learning techniques were used for data visualization, spatial analysis, and zonal statistics. Lagos Island, Nigeria was used as a case study to demonstrate how to obtain a more precise population estimate at the lowest administrative jurisdiction and eliminate ambiguity caused by antithetical parameters in the calculations. We also demonstrated how the framework can be used as a benchmark for estimating the carrying capacities of urban basic services like healthcare, housing, sanitary facilities, education, water etc. The proposed framework would help urban planners and government agencies to plan and manage cities better using more accurate data.

[1] University of Lagos, Nigeria

[2] The Hong Kong Polytechnic University

3. Instituto Nacional de Investigación y Capacitación de Telecomunicaciones (INICTEL-UNI)

Correspondence: Victor E. Irekponor, University of Maryland, college park, email: vireks@umd.edu

## 1. Introduction

The decennial census data in many countries or regions serve as the major source of demographic data; however, in the majority of least developed and developing nations, these data are out-of-date or unavailable due to the costs and technicalities of conducting a census (Catherine Linard 2012). As a result of this challenge, it becomes quite difficult to make effective and efficient plans with these census data. In certain instances, the accuracy of the population projections is also questionable for a variety of reasons, which includes the omission or undercounting of marginalized ethnic minorities, those living in informal settlements, and nomads. Other reasons include politicians' attempts to increase resource allocation by inflating population estimates and lengthy delays in the distribution of census data due to corruption (Wardrop, Jochem et al. 2018). Most population census figures are aggregated at the macro levels such as states, provinces, or regions, and lack spatial representations. Therefore, they are coarse and contain little to no information about the smaller geographic boundaries within their jurisdiction.

In Africa's most populous country; Nigeria, for instance, the last decennial census was conducted in 2006, after shifting the exercise three times over the past 16 years, primarily due to issues of funding, insurgency in the northern parts of the country, corruption, and improper management (Isah, Mohammad et al. 2019). In order to estimate the current population of the country or parts of the country, the previous outdated census figure is multiplied by a fixed growth rate and different variables, such as mortality, natality, migration, amongst others, which in turn leads to different contradictory figures by different authorities, and on the internet. Currently, the Lagos state government estimates its population to be 17.5 million people, while the National Population Commission (NPC) of the federal government estimates it to be 21.5 million, and the World Bank predicted in 2018 that the population of Lagos state would reach 25 million by 2020 (Lagos-state-govt 2019).

This ambiguity in population estimates has culminated in the use of state-of-the-art GIS techniques, remote sensing and machine learning to estimate population by motivated researchers, particularly in cases where inaccessibility or security concerns prevent the collection of census data, or where the most recent census conducted has become outdated (Wardrop, Jochem et al. 2018, Weber, Seaman et al. 2018). Given that the world population will increase by 2.3 billion people between 2011 and 2050, with more than 50 percent of that growth concentrated in urban areas alone (Leeson 2018), it is more than pertinent to comprehend the ramifications of not having reliable population figures for a geographic area. Fortunately, GRID3 (Geo-Referenced Infrastructure and Demographic Data for Development)[1] has already done an excellent job of resolving this ambiguity in population estimates by producing spatially disaggregated high-resolution population data estimates for all of Nigeria, given that the population census figures are outdated. In addition, the spatial disaggregation of the population dataset makes it possible to obtain population estimates at micro-levels by performing additional spatial analysis and raster calculations using QGIS and Python.

The GRID3 spatially disaggregated population data estimates for Nigeria were generated using a cutting-edge combination of GIS, remote sensing, and machine learning techniques. The Bottom-up modelling approach was adopted, which involves modelling population by combining geospatial covariates with a microcensus as inputs for a statistical model. A Bayesian Hierarchical Regression model within the family of poisson generalized linear mixed models is then used to model population estimates for unsampled regions. (Wardrop, Jochem et al. 2018, GRID3 High-res. population estimates 2019). Other GIS-based approaches for estimating population from literature includes the work of Dimitris et al. (2016) that employed a combination of remotely sensed covariates and statistical analysis methods to obtain population estimates with high spatial resolution (Dimitris and Petros 2016). And those of Qiu, Sridharan et al. (2010), Dong, Ramesh et al. (2010), Joseph, Wang et al. (2013) and Tomás, Fonseca et al. (2015) that used LiDAR and high resolution satellite images to determine population estimates.

[1] The GRID3 project is funded by the Bill and Melinda Gates foundation and United Kingdom's Department for International Development, and the data was generated by the WorldPop Research Group at the University of Southampton.

GRID3 collaborates with academic researchers and the private sector to design relevant and adaptable open-source geospatial solutions for many African nations based on their capacity and development requirements. In addition to Nigeria, the initiative is currently active in the following countries: Cabo Verde, Zambia, Togo, Equatorial Guinea, Mauritius, Cameroon, Burkina Faso, Rwanda, Chad, Botswana, Eritrea, Lesotho, Mali, Sudan, Ethiopia, Uganda, Liberia, Angola, Niger, Republic of Congo, Djibouti, Comoros, Reunion, Burundi, Guinea, Senegal, The Gambia. The plan is to scale even further in the next years which will make near-accurate spatially disaggregated population datasets available and accessible for a large number of sub-Saharan African developing nations. (Blair-Freese 2019). This paper demonstrates how to aggregate this dataset at the lowest administrative level, or “ward” in the case of Nigeria. A ward is a smaller geographic area within a Local Government geographical area (Moore 2015). Also, as long as the shapefile of the administrative boundary is available, the framework presented in this paper can be implemented in any country or region where the GRID3 dataset is accessible.

## 2. Materials and Methods

### 2.1. *Research Framework*

The framework in Fig. 1 was designed to show how micro-level population estimates are obtained from the GRID3 spatially disaggregated population estimates. From Fig. 1, the grided population estimates for Nigeria is obtained from the GRID3 official website for the Nigerian government[2], and the Lagos state grided population estimates are extracted, which are then loaded into QGIS alongside the Lagos state ward administrative boundary shapefile. Within QGIS, a lot of different spatial analyses are performed including raster to vector transformations, raster calculations, clipping by masks, zonal statistics etc. Section 4 provides more details on that in a step-by-step manner.

The final outputs of the QGIS analysis would be a CSV file, charts, and data visualizations of micro-level (ward) population estimates for the Lagos island local government in Lagos state,

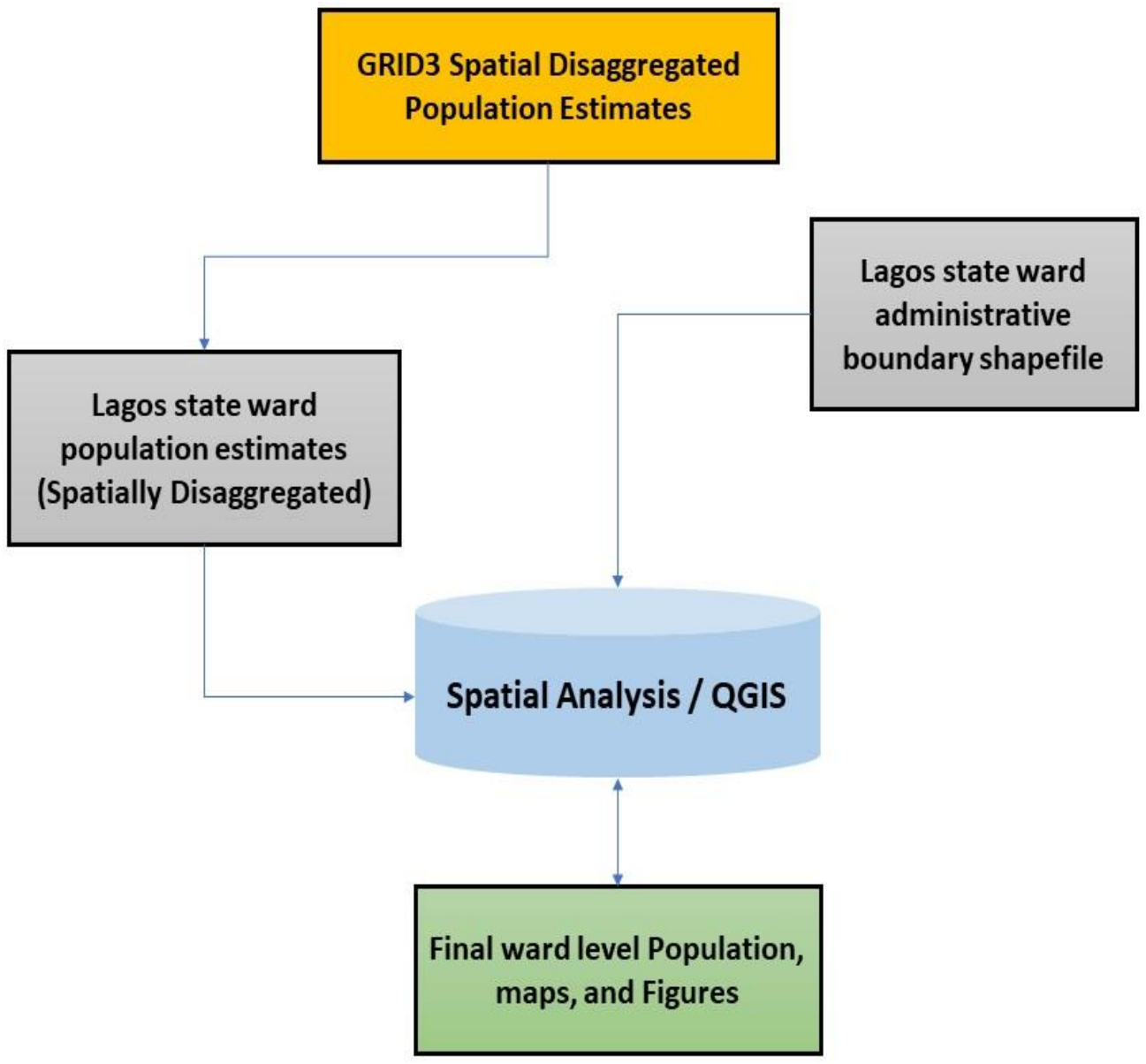

**Fig. 1** The Methodological architecture

[2] Grid3 official website for the Nigerian Government.
https://grid3.gov.ng/

Nigeria. If the GRID3 spatially disaggregated dataset is used, this methodology can be applied to and replicated for any administrative boundary.

2.2. *Source Data:*

Using this framework requires the GRID3 population data estimates and the local administrative boundary shapefile. For the purpose of this paper, the ward-level administrative shapefile for Lagos-island local government area in Lagos state, Nigeria was used, and at the core of the algorithm used to generate the GRID3 population data estimates is a Bayesian Hierarchical Poisson Regression model which may be specified as:

$$N_i \sim Poisson(D_i\ A_i) \qquad (1)$$

$$D_i \sim LogNormal(\mu_i, \sigma_{t,r,s,l}) \qquad (2)$$

$$\mu_i = \alpha_{t,r,s,l} + \sum_{k=1}^{k} \beta_k x_{k,i} \qquad (3)$$

Where $\boldsymbol{N_i}$ = Total population count, $\boldsymbol{D_i}$ = Population density (People/Hectare), $\boldsymbol{A_i}$ = Area of the settlement (Hectares), $\boldsymbol{x}$ = covariates, $\boldsymbol{i}$ = location, $\boldsymbol{\alpha}$ = intercept term of the regression, which is estimated for each settlement type ($\boldsymbol{t}$**),** region ($\boldsymbol{r}$), state ($\boldsymbol{s}$), and local government area ($\boldsymbol{l}$**)**, $\boldsymbol{\beta}$ = regression slope (the effect of $\boldsymbol{x}$ on $\boldsymbol{D}$), $\boldsymbol{\sigma}$ = unexplained variance, this means variation in $\boldsymbol{D}$ that cannot be explained by $\boldsymbol{x}$**.**

More specifically, a microcensus was collected between the years 2016 and 2017 in 1,141 clusters and in 15 states of the federation, and the total population count ($\boldsymbol{N}$**)**, Population density ($\boldsymbol{D}$**)**, and covariates ($\boldsymbol{x}$**)** were measured for each of the 1,141 microcensus locations (WorldPop and National Population Commission of Nigeria 2021). The information gotten was then used to estimate $\boldsymbol{\alpha}, \boldsymbol{\beta}$ and $\boldsymbol{\sigma}$**,** the covariates $\boldsymbol{(x)}$ was then measured for every location in the country. Important covariate datasets include the settlement type raster for Nigeria which was obtained from both LandScan (Oak Ridge National Laboratory, 2018) and Weber et al. (2018) map of settlement types, school density obtained from eHealth Africa[3], the WorldPop Global gridded population estimates (Stevens, Gaughan et al. 2015), and the Gridded household size data that was interpolated from the Demographic and Health Survey data (National Population Commission and ICF International, 2014).

The GRID3 population data estimates and shapefiles can be obtained from the official GRID3 website for the Nigerian government or from the global GRID3 data repository in the case of other African countries. The majority of the spatial analysis to obtain the ward-level population figures is performed within the QGIS software.

## 3. Results and Discussion

The following section shows the whole process as well as the steps for generating ward level population estimates for the wards in Lagos Island local government area:

1. Download the GRID3 population data estimates and the Lagos state ward level shapefile from the official website[4].
2. Load the vector and raster data to QGIS respectively using the Data Source Manager (DSM), pressing the ctrl + L key opens the DSM.
3. Next, extract the study area of Lagos Island LGA from the raster and vector data of the entire state. For the raster file, click on the raster-by-mask tool in QGIS (from the ribbon

[3] Derived from eHealth in Nigeria via http://geopode.world/
[4] https://grid3.gov.ng/dataset/national-population-estimates/resources

*raster>extraction>clip-raster-by-mask*). For the vector file, Lagos island LGA is selected from the attribute table and saved to a new file. This step gives the boundary and raster for just Lagos-island alone.

**4.** Tune the symbology elements in the layer properties. This includes setting the colour band for the raster, making the boundary layer transparent, and editing the layer information.

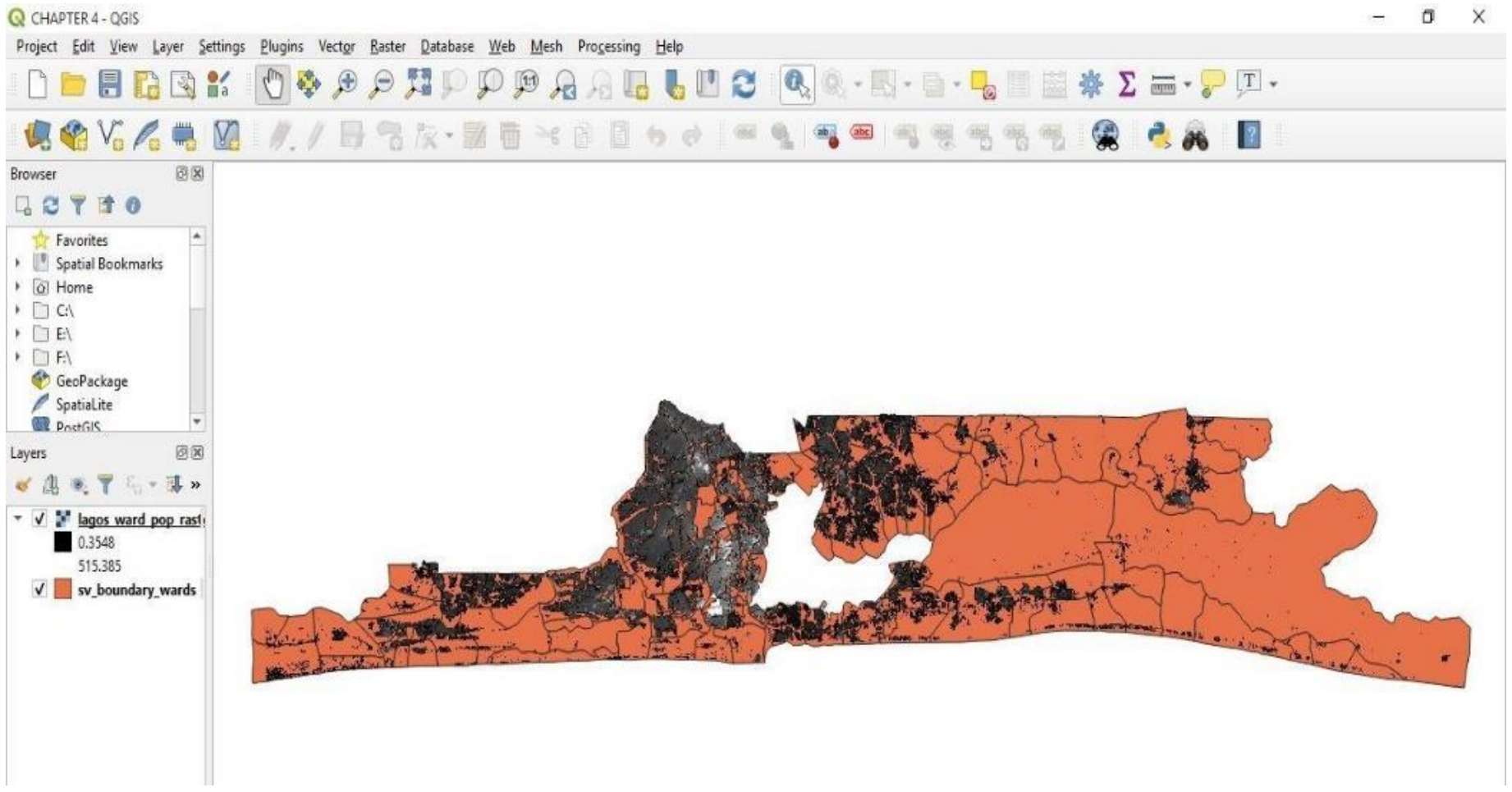

**Fig. 2** Showing the viewport of QGIS at step 3

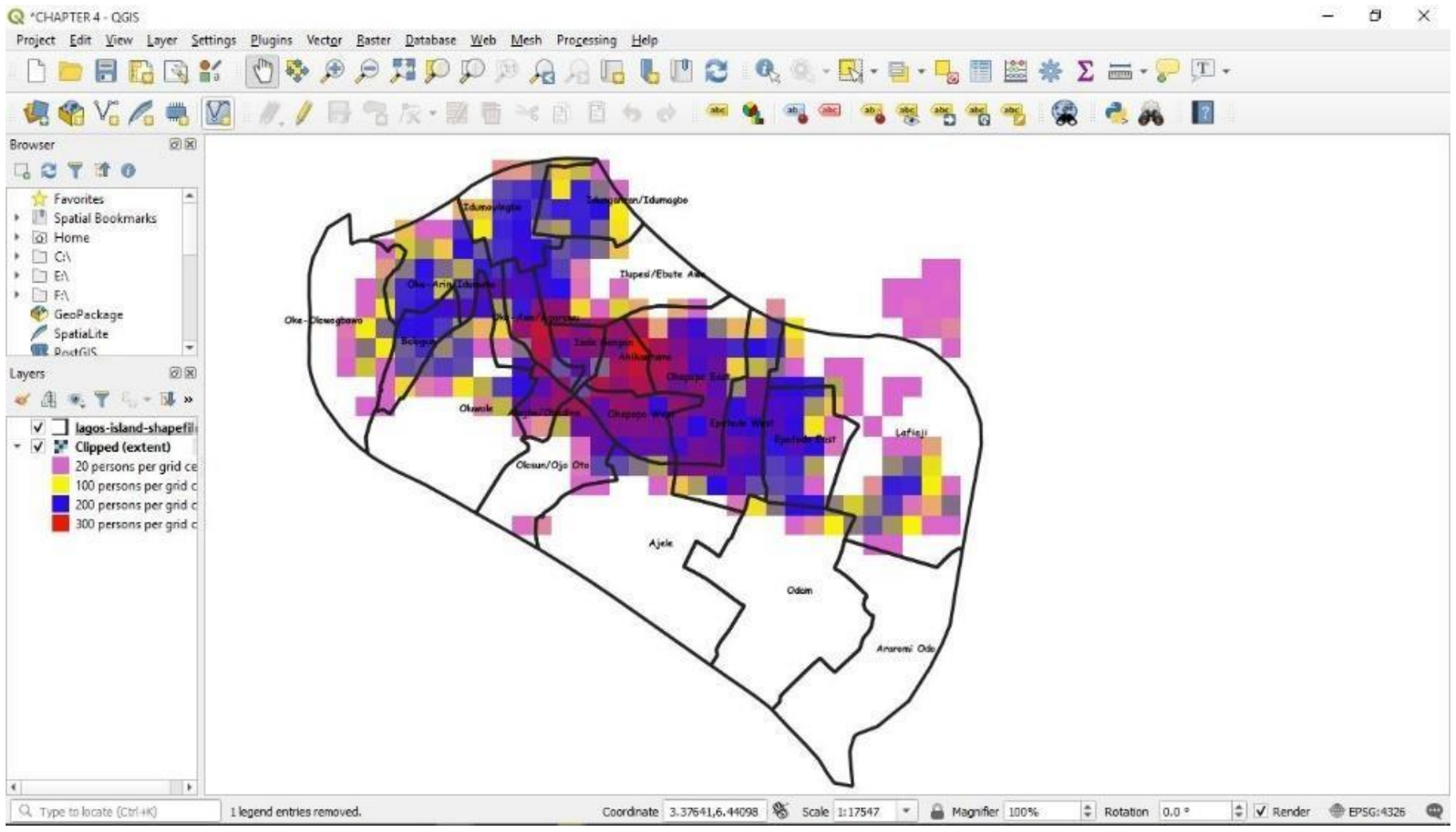

**Fig. 3** Showing the viewport of QGIS at step 4

**5.** Utilize the zonal statistics tool in QGIS to calculate the sum or mean-sum of each ward's population. This will enable us obtain population estimates at the ward level in Lagos-island

LGA. The zonal statistics tool can be located under *processing>toolbox>raster-analysis>zonal-statistics.*

**6.** To visualize the output of the zonal statistics algorithm, open the attribute table for the *Lagos-Island ward layer*. Three additional columns should be added to the existing columns by this algorithm namely:

i. _count

ii. _sum

iii. _mean

The _count column displays the computed number of grids inside a ward boundary, the _sum column displays the calculated number of persons the model estimates to be in a 100m x 100m grid, and the _mean column displays the average of all cells in the value raster that belong to the same zone as the output cell.

**7.** From the description in step 6, _sum is the column of interest for this study, as it outputs the estimated population figures for each ward in the study area. At this stage, the python console should be used to extract the three columns of interest; lga_name, ward_name, and _sum and save them in a single .csv file which can then be viewed in an Excel spreadsheet. The python script to generate and save the csv file of the final micro-level (ward) population estimates has been uploaded to the project's github and made open source; to use this, open the scripting shell in QGIS and copy-paste the code from the project's github[5].

As indicated by equations 1 through 3, the model used to generate the population estimate is a Bayesian framework, which is a probabilistic approach that provides estimates of the uncertainty

[5] url to project's github
https://github.com/marquisvictor/Micro-level-population-estimation/blob/main/step_7-script.py

in population estimates. Consequently, when producing estimated totals for micro-level areas such as wards, users of the framework proposed in this research and the GRID3 data should take cognizant of the fact that the uncertainty in estimates might be rather substantial in particular areas[6]. However, this is far better than guessing or multiplying an outdated population census figure with some arbitrary growth rate without considering immigration, emigration, mortality, or natality, as is the situation in Nigeria.

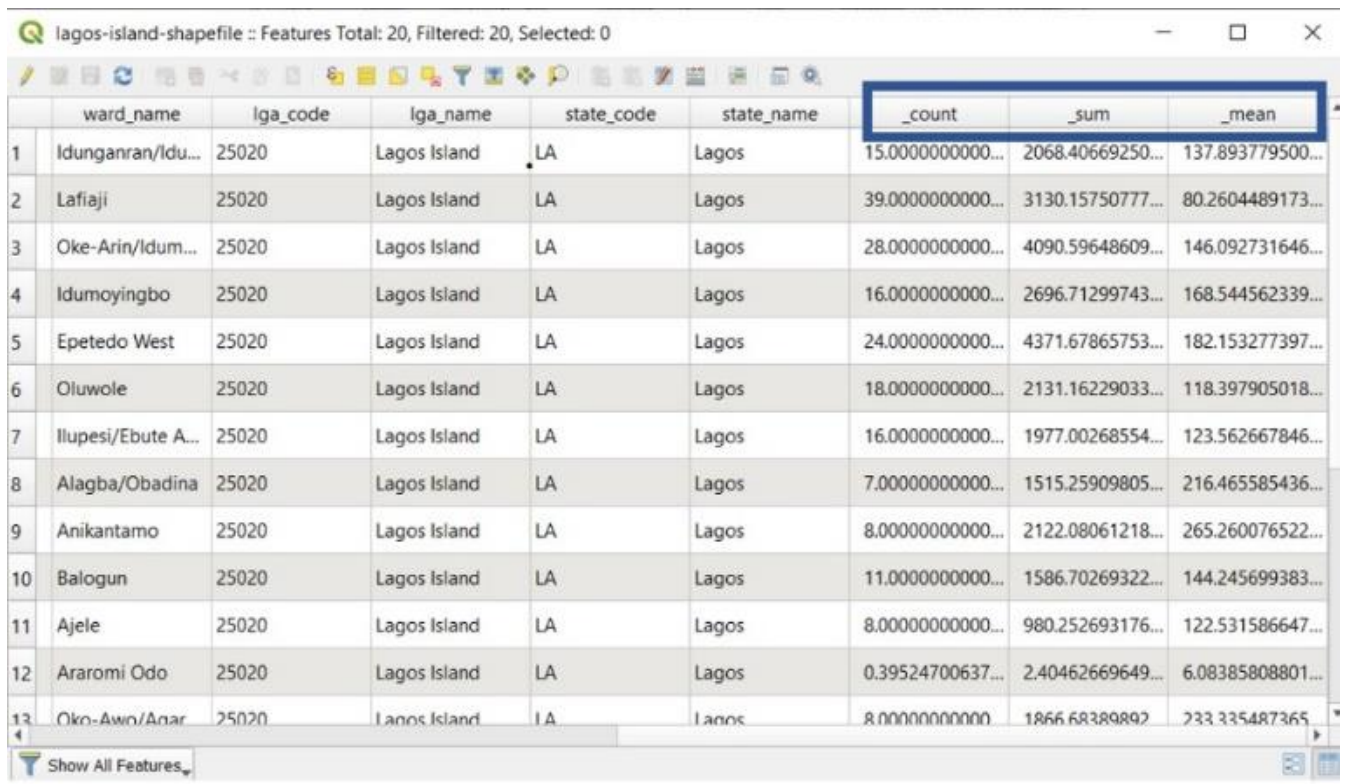
lagos-island-shapefile :: Features Total: 20, Filtered: 20, Selected: 0

| | ward_name | lga_code | lga_name | state_code | state_name | _count | _sum | _mean |
|---|---|---|---|---|---|---|---|---|
| 1 | Idunganran/Idu... | 25020 | Lagos Island | LA | Lagos | 15.0000000000... | 2068.40669250... | 137.893779500... |
| 2 | Lafiaji | 25020 | Lagos Island | LA | Lagos | 39.0000000000... | 3130.15750777... | 80.2604489173... |
| 3 | Oke-Arin/Idum... | 25020 | Lagos Island | LA | Lagos | 28.0000000000... | 4090.59648609... | 146.092731646... |
| 4 | Idumoyingbo | 25020 | Lagos Island | LA | Lagos | 16.0000000000... | 2696.71299743... | 168.544562339... |
| 5 | Epetedo West | 25020 | Lagos Island | LA | Lagos | 24.0000000000... | 4371.67865753... | 182.153277397... |
| 6 | Oluwole | 25020 | Lagos Island | LA | Lagos | 18.0000000000... | 2131.16229033... | 118.397905018... |
| 7 | Ilupesi/Ebute A... | 25020 | Lagos Island | LA | Lagos | 16.0000000000... | 1977.00268554... | 123.562667846... |
| 8 | Alagba/Obadina | 25020 | Lagos Island | LA | Lagos | 7.00000000000... | 1515.25909805... | 216.465585436... |
| 9 | Anikantamo | 25020 | Lagos Island | LA | Lagos | 8.00000000000... | 2122.08061218... | 265.260076522... |
| 10 | Balogun | 25020 | Lagos Island | LA | Lagos | 11.0000000000... | 1586.70269322... | 144.245699383... |
| 11 | Ajele | 25020 | Lagos Island | LA | Lagos | 8.00000000000... | 980.252693176... | 122.531586647... |
| 12 | Araromi Odo | 25020 | Lagos Island | LA | Lagos | 0.39524700637... | 2.40462669649... | 6.08385808801... |
| 13 | Oko-Awo/Agar... | 25020 | Lagos Island | LA | Lagos | 8.00000000000... | 1866.68389892... | 233.335487365... |

Show All Features

**Fig. 4** Showing the attribute table described in step 6

## 3.1. *Data Visualization*

The open-source python script prepared for this project automatically generates a bar chart showing the different wards in the local government as well as their estimated population figures, shown in Fig. 5.

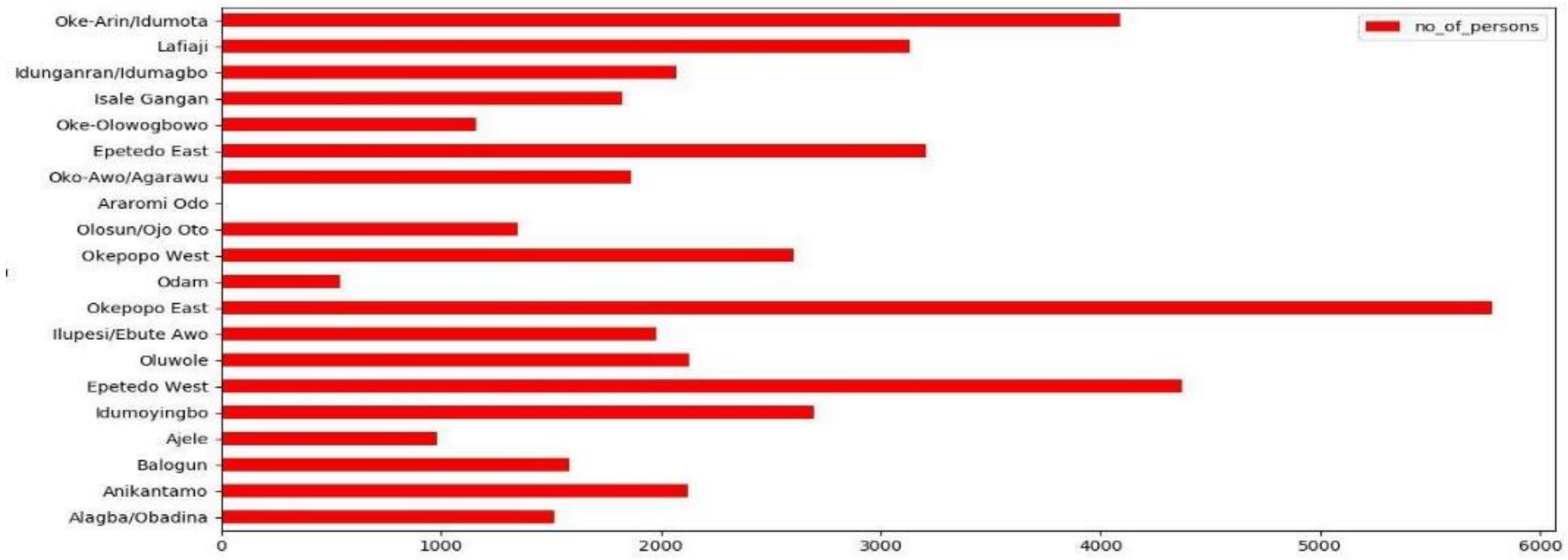

**Fig. 5** Bar chart of the different wards and their respective population counts

---

[6] understanding and considering the likely ranges of population for an area is critical. See interactive website https://apps.worldpop.org/woprVision/

Okepopo-east is the ward with the highest population in the local government, whereas Araromi-Odo has the lowest population. One can also view the breakdown of population estimates for each ward in the local government area. The national population census doesn't provide this level of granularity, that is, population estimates for each ward in the local government area; instead, only estimates at the local government level are available.

### 3.2. *Use cases*

For better appreciation of this framework, the following two use cases would be explored to validate the usefulness of this approach and to further demonstrate that it can be applied to any geography where GRID3 spatially disaggregated population estimates are available. These include i) Estimating affordable housing demand and supply in Lagos Island and, ii) determining the number of public toilets necessary to improve sanitation on Lagos Island.

#### 3.2.1. *Estimating housing demand and needs*

As a prerequisite to the survival of man, housing ranks second only to food (Enisan and Ogundiran 2013). Housing demand and housing need have different meanings and are most times confused for the other, it is worthy to note that every household has a need irrespective of the type of housing or income. Housing need has been defined by several researchers in the past; prominent among these definitions is Robinson's (1979), he defined housing need as the quantity of housing required for the accommodation of the agreed minimum standard and above for a population given its size and household composition without taking into account the household's ability to pay for the house assigned to it. This should not be confused with housing demand which is the relationship between the price of housing and the quantity and quality of housing for which people are willing and able to pay.

This study demonstrates how to obtain reliable population estimates at the smallest administrative level in the country; the ward level. This facilitates the decision makers' work and brings them one step closer to resolving several human and settlements challenges, such as housing demand (Paul Wallace 2019), and housing needs (Alao 2018).

According to the GRID3 model estimates, the nighttime residential population of Lagos-island local government area is around 74,000 people, with an average of approximately 2,300 people per ward. Having this information as a reference point or benchmark would go a long way toward resolving the housing problem in Lagos Island. Profiling the population, that is, gathering additional socioeconomic and demographic data on the population, such as household size, income level, occupation, and age ratio, would not be an arduous task. Once these data have been obtained, one can then determine the type and quantity of housing to be provided.

### 3.2.2. *Assessment of urban basic services e.g. public toilet needs in Lagos Island*

There is little proactive policy dimension for the provision and management of public toilets in Lagos island. Public toilets are spaces where particular salient needs can be fulfilled. Considering this, and the fact that public toilets are an integral part of public life in metropolitan cities such as the Lagos metropolitan area, their continuous availability and maintenance must always be ensured.

According to a study conducted by Olamiju et al. (2015), there is a dearth of public toilet facilities in Lagos island. In fact, a Google Maps search for "public toilets in Lagos island" return only two results: one at CMS inside Ajele ward and the other at Okepopo ward as shown in Fig. 6. In other words, there are no registered public toilets in the remaining 18 wards of a local government that servers over 45,000 people.

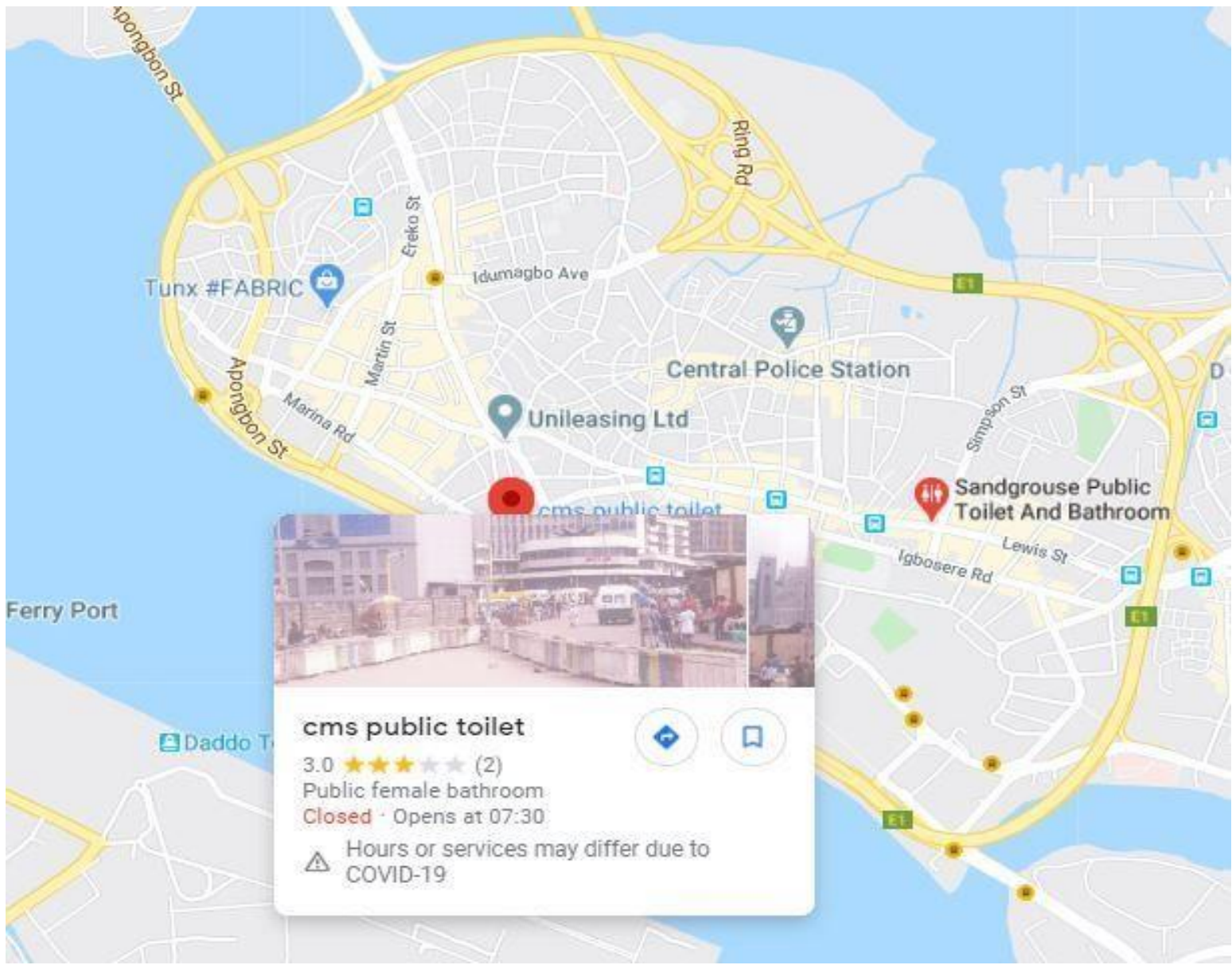

**Fig. 7** Google Maps search result for public toilets in Lagos Island

The Water, Sanitation, and Hygiene Foundation has revealed that Nigeria ranks second globally, after India, for the practice of open defecation (John-Mensah 2019). Consequently, this subject requires further attention. There should be a sufficient number of public toilets such that it should be commensurate with the total population within a given area's jurisdiction or radius coverage. in other words, there should be at least a set of public toilets in each ward in the local government area, but the number of public toilets should not be evenly distributed due to the uneven population distribution across wards. For instance, highly commercial wards, where you have the markets, such as; Balogun market, Idumota market, Jankara market, Oke arin market, etc., would generally have a higher daytime population and as such more public toilets should be provided in such areas than areas/wards that are entirely residential, the likes of Araromi odo, Odam, and Ajele.

According to British Standard 6465-1:2006+A1 (2009) – Sanitary Installations, male toilets need at least 4 water closets (WCs), 4 urinals, and 4 washbasins to accommodate 100 male users, while female toilets need at least 8 water closets and 8 washbasins to do the same (tradewashrooms 2020).

Using the state-of-the-art GRID3 population estimates, one can then extrapolate the total number of public toilets required across all wards in the study area based on these standards. The study's authors took the initiative to perform this extrapolation. Given that a toilet for a hundred male users must contain four WCs, four urinals, and four washbasins, and a toilet for a hundred female users must contain eight WCs and eight washbasins, python programming language and QGIS software were used to extrapolate and visually represent the number of male and female toilets required in each ward of the study area, given the relative reliability of their population estimates. The python script has been made open-source and uploaded to the project's github repository[7].

Fig. 7 depicts the dataframe object containing the generated estimates for toilets need across the wards in the study area, as well as the estimated amount of water closets, urinals, and washbasins for both the male and female categories, represented in the output dataframe by the *male units* and *female units'* columns

This analysis also reveals that Okepopo east, Idumota, Epetedo west, Epetedo east, and Idumoyingbo are the wards with the greatest need for additional public toilets due to their large population. As shown in Fig. 8, a bar-chart was plotted from the values in Fig. 7 to graphically visualize the total estimated toilet need for each ward.

[7] Link to the opensource code for the python script https://github.com/marquisvictor/Micro-level-population-estimation/blob/main/public_toilet-script.py

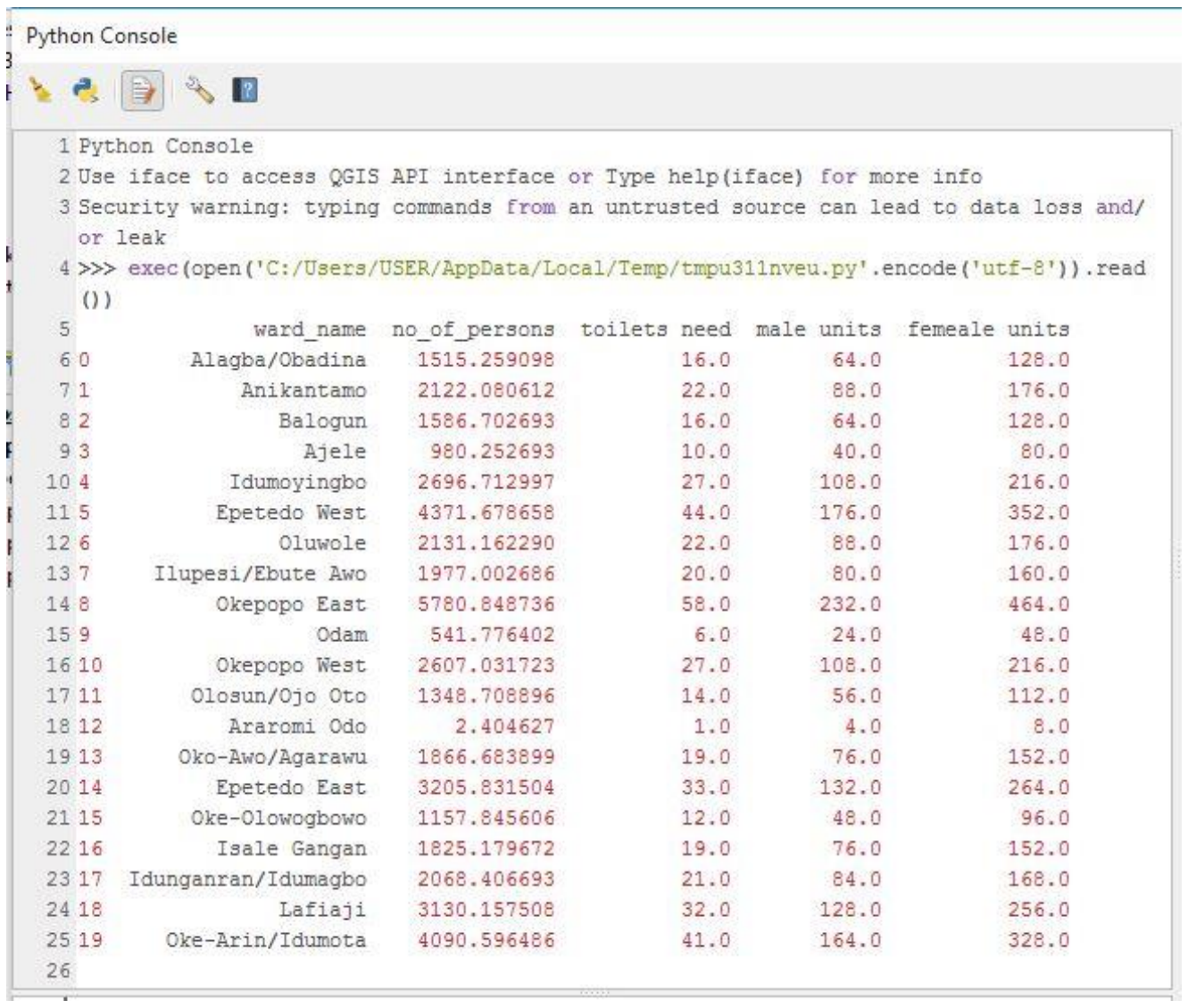

```
Python Console
Use iface to access QGIS API interface or Type help(iface) for more info
Security warning: typing commands from an untrusted source can lead to data loss and/
or leak
>>> exec(open('C:/Users/USER/AppData/Local/Temp/tmpu311nveu.py'.encode('utf-8')).read
())
             ward_name  no_of_persons  toilets need  male units  femeale units
0       Alagba/Obadina    1515.259098          16.0        64.0          128.0
1           Anikantamo    2122.080612          22.0        88.0          176.0
2              Balogun    1586.702693          16.0        64.0          128.0
3                Ajele     980.252693          10.0        40.0           80.0
4           Idumoyingbo   2696.712997          27.0       108.0          216.0
5         Epetedo West    4371.678658          44.0       176.0          352.0
6              Oluwole    2131.162290          22.0        88.0          176.0
7    Ilupesi/Ebute Awo    1977.002686          20.0        80.0          160.0
8         Okepopo East    5780.848736          58.0       232.0          464.0
9                 Odam     541.776402           6.0        24.0           48.0
10        Okepopo West    2607.031723          27.0       108.0          216.0
11      Olosun/Ojo Oto    1348.708896          14.0        56.0          112.0
12        Araromi Odo       2.404627           1.0         4.0            8.0
13     Oko-Awo/Agarawu    1866.683899          19.0        76.0          152.0
14        Epetedo East    3205.831504          33.0       132.0          264.0
15     Oke-Olowogbowo     1157.845606          12.0        48.0           96.0
16        Isale Gangan    1825.179672          19.0        76.0          152.0
17  Idunganran/Idumagbo   2068.406693          21.0        84.0          168.0
18             Lafiaji    3130.157508          32.0       128.0          256.0
19      Oke-Arin/Idumota  4090.596486          41.0       164.0          328.0
```

**Fig. 7** Dataframe containing the generated estimates for toilets need across the wards in the study area

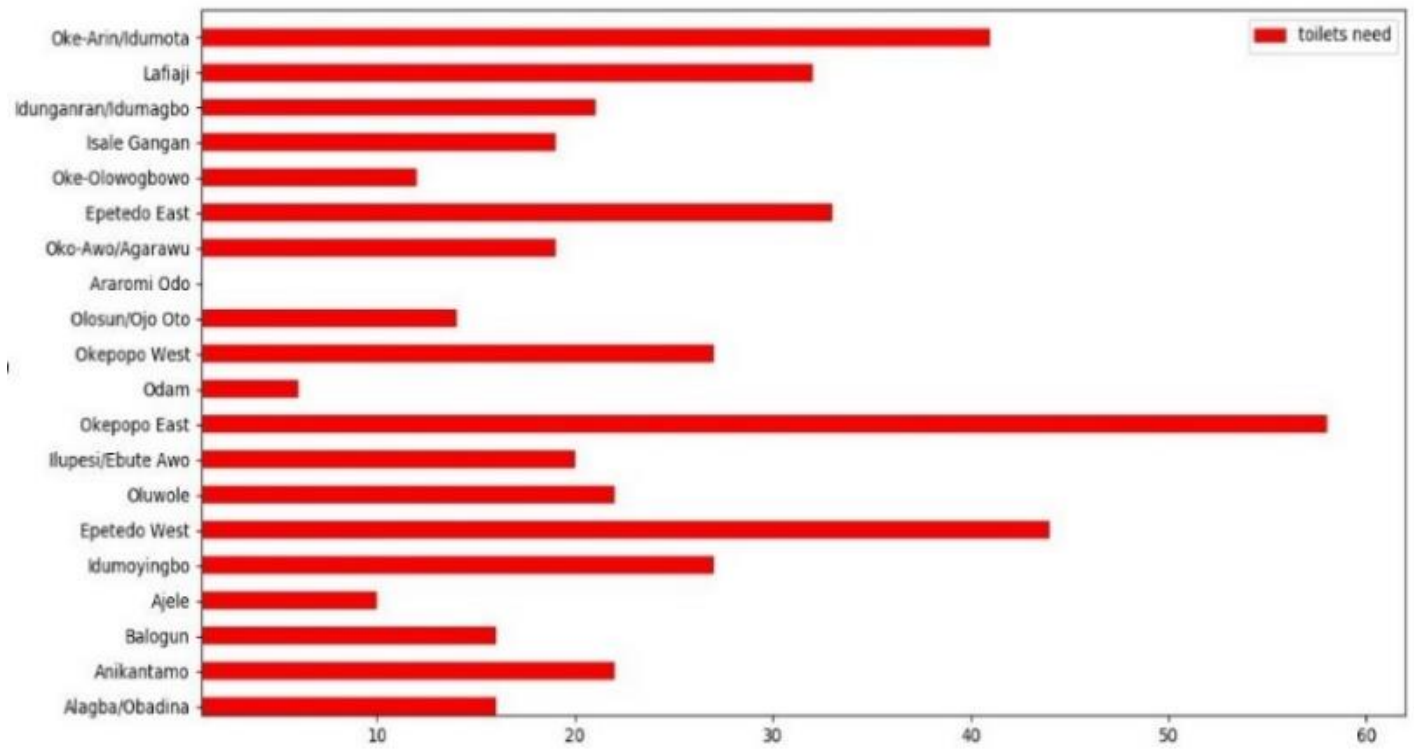

**Fig. 8** Bar-chart depicting the total estimated toilet need for each ward

## 4. Conclusion

This study demonstrates that near-accurate population estimates can be generated using state-of-the-art remote sensing techniques and aggregated at micro-levels. QGIS, Python programming language, and machine learning techniques were employed for spatial and raster analysis, data visualization, and zonal statistics respectively. More specifically, GRID3 population data estimates for Nigeria was used as the base dataset and the population estimates for Lagos state were extracted from it. Since the population estimates for the state of Lagos is spatially disaggregated, the population of Lagos island local government can further be extracted from it, as shown in section 4 of this article.

This framework allows us to obtain population estimates for each of the wards in Lagos island as ward level population estimates are not included in the GRID3 dataset repository; only state and local government-level estimates are available. Getting spatially disaggregated micro-level level population data estimates is very important for urban planners and policy makers to plan efficiently for the present and to project future demand for basic services. These include formulating policies, determining housing needs, and estimating the number of facilities and amenities needed, such as health care, transportation, schools, markets, among others. Considering that the last decennial census in Nigeria was conducted in 2006, the next source of near-accurate data for planning and management is the GRID3 population estimates for Nigeria, which used the bottom-up approach, that is, a combination of state-of-the-art remote sensing techniques and statistical modelling with geospatial covariates. It is important to note that this framework can be used to obtain population data estimates for any geographic boundary, given the availability of the shapefile boundary and GRID3 population data estimates. This methodology could be replicated in the sub-Saharan African regions of Chad, Benin, Côte d'Ivoire, Kenya, and Burundi, where the GRID3 program is currently active.

Lagos Island was used as a case study to demonstrate how to aggregate spatially disaggregated population estimates at the micro-level and resolve the ambiguity in population estimates arising from the use of antithetical parameters in the calculations. We also showed how the population data obtained can be used for estimating urban basic services such as healthcare, water, sanitary facilities, education, and housing supply. Essentially, the study provides urban planners, researchers and decision-makers access to high-quality population estimates which can be aggregated to any level of granularity.

**Conflict of Interest:** On behalf of all authors, the corresponding author states that there is no conflict of interest.

## References

Alao, T. (2018, Sep 12, 2018). "Lagos and the challenge of housing delivery." from https://tribuneonlineng.com/lagos-and-the-challenge-of-housing-delivery-2/. Blair-Freese, I. (2019). "Geo-Referenced Infrastructure and

Demographic Data for Development." IEEE Global Humanitarian Technology Conference (GHTC): 1. Catherine Linard, A. J. T. (2012). "Large-scale spatial population databases in infectious disease research." INTERNATIONAL JOURNAL OF HEALTH GEOGRAPHICS.

Dimitris, K. and P. Petros (2016). "Population Estimation in an Urban Area with Remote Sensing and

Geographical Information Systems." International Journal of Advanced Remote Sensing and GIS **5**(6): 1795-1812.

Dong, P., S. Ramesh and A. Nepali (2010). "Evaluation of small-area population estimation using LiDAR, Landsat TM and parcel data." International Journal of Remote Sensing **31**(21): 5571-5586.

Enisan, O. and A. Ogundiran (2013). "Challenges of Housing Delivery in Metropolitan Lagos." Research on Humanities and Social Sciences **Vol.3, No.20, 2013**.

GRID3 High-res. population estimates. (2019). "High-resolution population estimates." Retrieved 26-032021, 2021, from https://grid3.org/solution/high-resolution-population-estimates.

Isah, M., O. Mohammad and O. Nazariah (2019). "Nigerian National Population and Housing Census and Sustainable Development: Issues at Stake." Nigerian National Population and Housing Census and Sustainable Development: Issues at Stake **1**: 14.

John-Mensah, O. (2019, 26/06/2019). "Open Defecation - Nigeria Second Globally." from https://www.msn.com/en-xl/africa/nigeria/open-defecation-nigeria-second-globally/ar-AADrSbZ.

Joseph, M., L. Wang and F. Wang (2013). "Using Landsat Imagery and Census Data for Urban Population Density Modeling in Port-au-Prince, Haiti." GIScience & Remote Sensing **49**(2): 228-250.

Lagos-state-govt. (2019, 2022). "About Lagos." 2022, from https://hos.lagosstate.gov.ng/about-lagos/.

Leeson, G. W. (2018). "The Growth, Ageing and Urbanisation of our World." Journal of Population Ageing **11**(2): 107-115.

Moore, A. (2015). Spatial Disaggregation Techniques for Visualizing and Evaluating Map Unit Composition. United States Department of Agriculture Natural Resources Conservation Service**:** 28.

Paul Wallace, T. A. (2019, December 20, 2019, 7:00 AM GMT+1). "Lagos Building Luxury Homes in Face of Affordable Housing Crisis." Retrieved December 20, 2019, 7:00 AM GMT+1, 2019, from https://www.bloomberg.com/news/articles/2019-12-20/lagos-building-luxury-homes-in-face-ofaffordable-housing-crisis.

Qiu, F., H. Sridharan and Y. Chun (2010). "Spatial Autoregressive Model for Population Estimation at the Census Block Level Using LIDAR-derived Building Volume Information." Cartography and Geographic Information Science **37**(3): 239-257.

Stevens, F. R., A. E. Gaughan, C. Linard and A. J. Tatem (2015). "Disaggregating census data for population mapping using random forests with remotely-sensed and ancillary data." PLoS One **10**(2): e0107042.

Tomás, L., L. Fonseca, C. Almeida, F. Leonardi and M. Pereira (2015). "Urban population estimation based on residential buildings volume using IKONOS-2 images and lidar data." International Journal of Remote Sensing **37**(sup1): 1-28.

tradewashrooms. (2020). "How Many Toilets Do You Need?", from https://www.tradewashrooms.co.uk/washroom-guidance/how-many-toilets-do-you-need/.

Wardrop, N. A., W. C. Jochem, T. J. Bird, H. R. Chamberlain, D. Clarke, D. Kerr, L. Bengtsson, S. Juran, V. Seaman and A. J. Tatem (2018). "Spatially disaggregated population estimates in the absence of national population and housing census data." Proc Natl Acad Sci U S A **115**(14): 3529-3537.

Weber, E. M., V. Y. Seaman, R. N. Stewart, T. J. Bird, A. J. Tatem, J. J. McKee, B. L. Bhaduri, J. J. Moehl and A. E. Reith (2018). "Census-independent population mapping in northern Nigeria." Remote Sens Environ **204**: 786-798.

WorldPop and National Population Commission of Nigeria (2021). "Bottom-up gridded population estimates for Nigeria, version 2.0." WorldPop University of Southampton: 9.